\documentstyle[epsf,epsfig,11pt]{article}
\begin{document}

\renewcommand{\arraystretch} {1.3}

\def\NPB#1#2#3{{\rm Nucl.~Phys.} {\bf{B#1}} (#2) #3}
\def\PLB#1#2#3{{\rm Phys.~Lett.} {\bf{B#1}} (#2) #3}
\def\PRD#1#2#3{{\rm Phys.~Rev.} {\bf{D#1}} (#2) #3}
\def\PRL#1#2#3{{\rm Phys.~Rev.~Lett.} {\bf{#1}} (#2) #3}
\def\EPJ#1#2#3{{\rm Eur.~Phys.~J.} {\bf C#1} (#2) #3}
\def\CPC#1#2#3{{\rm Comp.~Phys.~Comm.} {\bf#1} (#2) #3}
\def\JHEP#1#2#3{{\rm JHEP} { \bf{#1}} (#2) #3}
\newcommand{\etal}{{\it et al.}}

\def\pythia8{\ifmmode {\mbox{\textsc{Pythia8}}}\else
                 {\textsc{Pythia8}}\fi}%

\def\qqUg{\ifmmode {q\bar{q} \rightarrow \unpU / G + g}\else
                           {$q\bar{q} \rightarrow \unpU / G + g$}\fi}%
\def\qgUq{\ifmmode {qg \rightarrow \unpU / G + q}\else
                           {$qg \rightarrow \unpU / G + q$}\fi}%
\def\ggUg{\ifmmode {gg \rightarrow \unpU / G + g}\else
                           {$gg \rightarrow \unpU / G + g$}\fi}%

\def\ffUZ{\ifmmode {f\bar{f} \rightarrow \unpU / G + Z}\else
                           {$f\bar{f} \rightarrow \unpU / G + Z$}\fi}%
\def\ffUgam{\ifmmode {f\bar{f} \rightarrow \unpU / G + \gamma}\else
                           {$f\bar{f} \rightarrow \unpU / G + \gamma$}\fi}%

\def\ffgg{\ifmmode {f\bar{f} \rightarrow \unpU^* / G^* \rightarrow \gamma \gamma}\else
                           {$f\bar{f} \rightarrow \unpU^* / G^* \rightarrow \gamma \gamma$}\fi}%
\def\gggamgam{\ifmmode {gg \rightarrow \unpU^* / G^* \rightarrow \gamma \gamma}\else
                           {$gg \rightarrow \unpU^* / G^* \rightarrow \gamma \gamma$}\fi}%
\def\ffll{\ifmmode {f\bar{f} \rightarrow \unpU^* / G^* \rightarrow \ell \bar{\ell}}\else
                           {$f\bar{f} \rightarrow \unpU^* / G^* \rightarrow \ell \bar{\ell}$}\fi}%
\def\ggll{\ifmmode {gg \rightarrow \unpU^* / G^* \rightarrow \ell \bar{\ell}}\else
                           {$gg \rightarrow \unpU^* / G^* \rightarrow \ell \bar{\ell}$}\fi}%

\def\Uj{\ifmmode {\unpU / G + jet}\else
                           {$\unpU/G+jet$}\fi}%
\def\UgZ{\ifmmode {\unpU / G + Z/\gamma}\else
                           {$\unpU / G + Z/\gamma$}\fi}%
\def\gg{\ifmmode {\gamma \gamma}\else
                           {$\gamma \gamma$}\fi}%
\def\ll{\ifmmode {\ell \bar{\ell}}\else
                           {$\ell \bar{\ell}$}\fi}%

\def\unpU{\ifmmode {\cal U}\else
                 {$\cal U$}\fi}%


\begin{titlepage}

\pagenumbering{arabic}

\begin{flushright}
       MAN/HEP/2009/20 \\
       MCnet/09/20 \\
       DESY 09-214 \\
       Dec 2009 \\
\end{flushright}
\vspace*{0.5cm}
\begin{center}
\boldmath
{\Large \bf Real Emission and Virtual Exchange of Gravitons and Unparticles in Pythia8 } \\
\unboldmath
\end{center}
\vspace*{2.0cm}
{\large \bf 
S.~Ask$^{1*}$, 
I.~V.~Akin$^{2}$,
L.~Benucci$^{3}$,
A.~De~Roeck$^{3,4}$,
M.~Goebel$^{5,6}$,
J.~Haller$^{6}$
} \\
{\it 1) University of Manchester, UK.} \\
{\it 2) Middle East Technical University, Ankara, Turkey.} \\
{\it 3) Universiteit Antwerpen, Belgium.} \\
{\it 4) CERN, Geneva, Switzerland.} \\
{\it 5) DESY, Hamburg, Germany.} \\
{\it 6) Universit\"at Hamburg, Germany.} \\
{\tt * E-mail: Stefan.Ask@manchester.ac.uk }

\noindent \rule{\textwidth}{0.5ex}

\vspace*{2.0cm}

\begin{abstract}
\noindent
Models with large extra dimensions as well as unparticle models could 
give rise to new phenomena at collider experiments due to 
real emission or virtual exchange of gravitons or unparticles. In this 
paper we present the common implementation of these processes in the 
Monte Carlo generator \pythia8, using relations between the parameters 
of the two models. The program offers several options related to the 
treatment of the UV region of the effective theories, including the 
possibility of using a form factor for the running gravitational coupling. 
Characteristic results obtained with \pythia8 have been used to validate 
the implementations as well as to illustrate the key features and effects 
of the model parameters. The results presented in this paper are focused 
on mono-jet, di-photon and di-lepton final states at the LHC.

\end{abstract}

\vspace{2.cm}

\vspace{\fill}
\end{titlepage}

\pagebreak

\setcounter{page}{1}


\section{Introduction}
Models with large extra dimensions (LED) are popular extensions of the 
Standard Model (SM) and studies of LED phenomena are usually based on 
the so-called ADD scenario~\cite{bib:add98}, where gravity alone has 
access to the extra dimensions. The large size of the extra dimensions 
gives rise to dense Kaluza-Klein (KK) mass modes which appear as a 
continuous graviton mass spectrum. Due to the large number of 
contributing mass states graviton processes could obtain sufficiently 
large cross sections to be observed at collider experiments, where the 
new processes involve real emission or virtual exchange of gravitons. 
These processes have been studied in great detail \cite{bib:led08}, 
however, no signs of new physics have been indicated at existing 
experiments.

More recently so-called unparticle models~\cite{bib:geo07} have also 
gained a large interest. These models relate to physics originating 
from a scale invariant new sector which is coupled to the SM through 
a connector sector with a high mass scale. Unparticle models can often, 
at least from a phenomenological point of view, be associated with 
extra dimension models and the particular case addressed here would 
result in phenomena very similar to the scenario of LED gravity. 
Having access to processes with the same final state from more than 
one model can be helpful to generalise the experimental analysis as 
well as for studying differences in the measured observables. The 
latter, in turn, could also be of value after a discovery to 
characterise the observed signal. 
 
This paper describes the processes involving real emission or virtual 
exchange of unparticles ($\unpU$) or LED gravitons ($G$) as implemented 
in \pythia8\ \cite{bib:pythia8}. The corresponding processes are:
\begin{itemize}
\item{Mono-jets (\Uj ): \ggUg ,~ \qgUq , ~\qqUg  ;}
\item{Mono-$Z / \gamma$ (\UgZ ): \ffUZ ,~ \ffUgam ;}
\item{Di-photons (\gg ): \ffgg ,~ \gggamgam ;}
\item{Di-leptons (\ll ): \ffll ,~ \ggll .} 
\end{itemize}

Because of the large similarities both models can be covered by a 
common implementation \cite{bib:ask09}, where only a simple translation 
of a few model related constants is required to change between the 
models. In addition to technical advantages, this common description 
makes comparisons between the processes both consistent and transparent. 
The effective theory description of these processes behaves poorly in 
the UV limit and several related options are available in the program. 
These include a form factor \cite{bib:hew07} for the graviton processes which represents 
a realistic alternative of the behaviour at high energies due to the 
running of the gravitational coupling. 

The \pythia8 implementations have been validated against the original 
theory papers and we present here some of these results. Besides 
providing reference results for the \pythia8\ program, the results 
illustrate different characteristic effects of the model parameters 
and various key features\footnote{In order to prevent clutter, the 
figures presented in this paper do not include statistical errors, 
which should not be significant for the conclusions.}. All results 
are produced with respect to the nominal LHC conditions ($pp$-collisions 
at a centre-of-mass energy of $\sqrt{s} = 14$ TeV), unless explicitly 
stated otherwise. 

This paper is organised as follows. The first section starts with a 
description of the model parameters and different conventions. It is 
followed by a description of the implementation of the processes and 
the relations between the two models. This section also describes the 
available options for treating the UV region of the effective theories. 
Sec.~\ref{sec:remj} to \ref{sec:velp} present the different program 
options in \pythia8\ together with results related to the individual 
processes: \Uj , \gg , and \ll\ production. In addition, Sec.~\ref{sec:velp} 
draws special attention to possibile effects on the forward-backward 
asymmetry in lepton pair production from spin-1 unparticle exchange.
The \UgZ\ processes are not addressed here since they were already 
presented in a previous paper \cite{bib:ask09}. A complete list of 
the process and model parameter names used in the \pythia8\ program 
can be found in the appendix. The appendix also contains complementary 
information used for the scalar unparticle process, $f \bar{f} \rightarrow 
Z \unpU $, which is not available in the literature.

This paper presents results obtained with \textsc{Pythia} version 8.125 
and all parameters not explicitly mentioned in the text have been kept 
at the default values of this version. During the work a few minor 
corrections have been made\footnote{The corrections are related to the 
\ll\ process and, in addition, the spin-0 case of the $\unpU + Z/\gamma$ 
process has been added.}, which all are included in the following version 
8.130. In order to produce results comparable to what is available in the 
literature, only the hard part of the process has been considered\footnote{Where 
the program parameters {\tt PartonLevel:MI}, {\tt PartonLevel:ISR} and 
{\tt PartonLevel:FSR} have been turned off.} and the MRST2001 parton 
distribution functions (PDF) of the proton have been used unless stated 
otherwise.


\section{Implementation in \pythia8 \label{sec:ip}}
The common implementation of the LED graviton and unparticle processes 
in \pythia8\ is generally based on the cross section expressions derived 
for the unparticle case. These expressions can be converted into the 
corresponding LED graviton formulae by simple translations of a few model 
related constants. For this reason we first discuss the unparticle process 
and afterwards we introduce the translation into its LED equivalent. The 
values of the model parameters are sometimes constrained by arguments 
valid in different scenarios of the model, however, in order to allow for 
more phenomenologically motivated studies the implementations in \pythia8\ 
generally allow for any parameter value that is technically possible. 
Particular model constraints therefore have to be respected explicitly 
by the users. As mentioned above, the common implementation of the \UgZ\ 
processes was originally presented in \cite{bib:ask09}, together with a 
study of $U/G + Z$ production at the LHC, but for completeness relevant 
parts related to these processes will be repeated here.

\subsection{Parameter Conventions \label{sec:pc}}
The parameter convention used here for the unparticle case follows \cite{bib:che07}. 
It is assumed that any potential gaps in the unparticle mass spectrum 
are sufficiently small not to affect the high energy behaviour at particle 
colliders and that the life time is long enough to prevent decays at 
distances relevant for collider experiments. The main parameters are a 
scale dimension parameter ($d_\unpU$), an unparticle renormalisation scale 
($\Lambda_\unpU$) and a coupling constant ($\lambda$) of the unparticles to 
the SM fields, which is related to the mass scale of the connector sector.

In principle there are restrictions on the allowed ranges of the parameters. 
Full conformal invariance restricts the scaling dimension of the unparticle 
to $d_\unpU > 1$ (scalar), $d_\unpU > 3$ (vector) and $d_\unpU > 4$ (tensor), 
which follow from Mack's unitary constraint, as discussed for example in 
\cite{bib:geo09}. By assuming scale invariance, but not full conformal 
invariance, these restrictions are relaxed. In this paper we take a 
phenomenological approach and allow for the full $d_\unpU$ range and the 
lowest bound $d_\unpU > 1$ is used as the overall lower limit in \pythia8. 
For some processes it is, however, necessary to constrain this parameter to 
the range $1 < d_\unpU < 2$ for numerical reasons.  For $d_\unpU > 2$, 
especially in the case of virtual unparticle exchange, contributions that 
depend on the UV completion of the theory are relevant which suggests that 
the effective theory used will not be valid. The ranges allowed in \pythia8\ 
are in accordance with \cite{bib:che07,bib:kum08}, where for example Eq. (4) 
in the latter paper shows that the propagator diverges\footnote{The relation 
between $d_\unpU$ and the UV sensitivity has for example been studied in 
\cite{bib:cac09}.} when $d_\unpU \rightarrow 2$, 
\begin{eqnarray}
\int d^4xe^{iPx} \langle 0 | T O^{\kappa}_{\cal U}(x) O^{\kappa}_{\cal U}(0) | 0 \rangle \propto 
\frac{A_{d_\unpU}}{2 \sin{(d_\unpU\pi)}} = \chi_\unpU \\
A_{d_\unpU} = \frac{16\pi ^2 \sqrt{\pi}}{(2\pi)^{2d_\unpU}}\frac{\Gamma(d_\unpU + \frac{1}{2})}{\Gamma(d_\unpU - 1)\Gamma(2d_\unpU)} 
\end{eqnarray}
where $ A_{d_\unpU}$ is a normalisation constant related to the rather 
peculiar unparticle phase space. Hence, this divergence implies that the virtual 
\unpU\ exchange processes in \pythia8\ require $d_{U} < 2$. The formulae 
obtained for the \unpU\ emission processes on the other hand give finite 
results also for the range $d_\unpU > 2$. However, in this case the results 
have to be interpreted more carefully as discussed in Sec. \ref{sec:tethe}.

The parameter convention used here for LED gravity follows \cite{bib:giu99}. 
The gaps between the individual KK modes of the graviton are determined 
by the size of the extra dimensions together with a possible curvature 
of space-time, like in the so-called Randall-Sundrum scenarios~\cite{bib:rs99}. 
It has been shown that a small curvature can remove cosmological constraints 
from low energy processes~\cite{bib:giu05}. However, for the processes 
addressed here it is assumed that any mass gaps in the graviton mass 
spectrum are sufficiently small for the high energy phenomena to be 
unaffected. 

The main model parameters correspond to the fundamental scale of $D ~(= n + 4)$ 
dimensional gravity ($M_D$), the number of extra dimensions ($n$) and in 
the case of virtual graviton exchange an effective cut-off scale ($\Lambda_T$). 
The extra dimensions are normally assumed to form a $n$-dimensional torus 
with a common radius ($R_D$). This radius is determined by $n$, $M_D$ 
together with the 4-dimensional Planck mass and is for this reason not 
included as a parameter for the processes in \pythia8. A number of different 
conventions are used in the literature and the most popular are related as 
follows:
\begin{itemize}
\item{GRW \cite{bib:giu99}: $M_D$, $R_D$, $n$, $\Lambda_T$ (used in \pythia8);}
\item{MPP \cite{bib:mir99}: $M^{n+2} = 2 M_D^{n+2}$;}
\item{HLZ \cite{bib:han98}: $M^{n+2}_S = 8 \pi ^{1-\frac{n}{2}}~\Gamma(\frac{n}{2})M_D^{n+2}$, $R_S = 2 \pi R_D$;}
\item{HR  \cite{bib:hew07}: $\Lambda^4_H = \frac{2}{\pi} \Lambda_T^4$.}
\end{itemize}

\subsection{Implementation of the Processes \label{sec:iotp}}
The processes in \pythia8\ are available for unparticles with spin 0, 1 and 2. 
However, not all spin values are available for all processes. The \UgZ\ 
processes are implemented as 2--to--2 parton level processes based on the $Z$ 
matrix elements (ME) calculated in \cite{bib:che07} for spin-1 and spin-2 
unparticle emission. In addition the spin-0 case is available, which is based 
on the ME given in the appendix. In the case of $Z$ production, no $Z/\gamma^*$ 
interference is included and the $Z$ decays isotropically. The photon process 
corresponds to the photon limit of the $Z$ process. All $\unpU/G$ emission 
processes implemented in \pythia8\ share the graviton particle code 5000039, 
since the $\unpU$ or $G$ has a continuous mass spectrum. This is in contrast 
to processes related to an individual mass state, where for example the first 
KK mode of the graviton would have the particle code 5100039. The main difference 
with respect to existing processes in \pythia8\ was the continuous mass spectrum 
of the unparticle. This has been solved by generating unparticle masses, using 
already existing \pythia8\ functionality, according to a Breit-Wigner distribution. 
The events are then re-weighted in order to obtain a correct mass distribution, 
as given by the differential cross section. For this reason the Breit-Wigner 
shape should be matched to the relevant range of the mass spectrum in order to 
achieve the highest possible Monte Carlo (MC) efficiency and generation speed. 
The Breit-Wigner shape is defined by the program parameters, {\tt 5000039:m0, 
5000039:mWidth, 5000039:mMin, 5000039:mMax}, according to the standard particle 
data scheme in \pythia8.

The \Uj\ processes have been implemented in the same way as the \UgZ\ processes. 
They are available for spin-0 and spin-1 unparticles, however, graviton emission 
is the only available spin-2 scenario. The processes are based on MEs from 
\cite{bib:giu99} for $G$ emission and \cite{bib:che07,bib:riz08} for the 
$\unpU$ processes. The spin-0 unparticle interactions used for the processes 
presented here are given by the effective operators \cite{bib:che07},
\begin{equation}
{\cal L} \supset \frac{\lambda}{\Lambda_\unpU^{d_\unpU - 1}}\sum_{f} \bar{f}fO_\unpU 
+ \frac{\lambda}{\Lambda_\unpU^{d_\unpU}}\sum_{G}G_{ \alpha \beta }G^{\alpha \beta }O_\unpU
\label{eq:unpscalar}
\end{equation} 
where $f$ refers to fermions and $G_{\alpha\beta}$ to gauge bosons. Both terms 
in Eq. (\ref{eq:unpscalar}) lead to amplitudes at leading order for, \qqUg\ 
and \qgUq , whereas \ggUg\ only involves the gluon vertex, {\it i.e.} the second 
term. Since the gluon vertex is suppressed by one additional power of $\Lambda_\unpU$ 
compared to the fermion vertices, {\it i.e.} two powers with respect to the 
cross section, the MEs of the two quark initiated processes are approximated 
by including only diagrams with unparticle emission from a fermion line. The 
coupling constants of the two terms are not necessarily the same and it could 
be of interest to study them separately. Again due to the suppression by different 
powers of $\Lambda_\unpU$ in the two terms, this can be approximated by using 
\qqUg\ and \qgUq\ to represent the contributions from the quark vertices and 
\ggUg\ to represent the gluon vertex contribution.

In order to obtain the corresponding graviton emission processes, the 
unparticle parameters have been interpreted as follows,
\begin{eqnarray}
r = \lambda'_2 / \lambda_2 = 1 \\
d_\unpU = \frac{n}{2} + 1 \\
\Lambda_\unpU = M_D \\
A_{d_\unpU} \rightarrow S'(n) = \frac{2\pi ~ \pi^{n/2}}{\Gamma(\frac{n}{2})}
\end{eqnarray}
where $\lambda_2$ and $\lambda'_2$ refer to the possibility of having 
different couplings related to the two effective spin-2 unparticle operators 
\cite{bib:che07}. This translation is automatically done inside \pythia8\ 
when the graviton processes are used.

The virtual $\unpU/G$ exchange processes are implemented in \pythia8\ as 
2--to--2 parton level processes. Certain processes contain coherent SM 
amplitudes and converge to the SM results if the unparticle contribution 
is close to zero, {\it e.g.} by setting $\lambda = 0$ or $\Lambda_\unpU \rightarrow \infty$. 
The SM results shown for comparison in the figures below have been obtained 
by the corresponding $\unpU/G$ process using such extreme parameter values.

The \gg\ processes are available for spin-0 and spin-2 unparticles based 
on the matrix elements in~\cite{bib:che07,bib:kum08}. In the spin-2 case, 
the fermion initiated process includes the corresponding SM $t$-channel 
process (equivalent to {\tt PromptPhoton:ffbar2gammagamma}) together with 
the related interference. The spin-2 gluon process does not involve the 
possible SM box diagram which for this reason must be included separately, 
{\it e.g.} using the \pythia8\ process\footnote{Note that this process is 
based on a massless approximation which can be relevant.} {\tt PromptPhoton:gg2gammagamma}.

The \ll\ processes only concern charged lepton pair production. The fermion 
initiated process is available for both spin-1 and spin-2 unparticles and 
includes the SM $Z/\gamma^*$ amplitudes and interference. The gluon process 
is only available for spin-2 and does not contain any SM contributions. All 
processes are implemented following the MEs in \cite{bib:che07}. Since the 
fermion initiated process could be used for studies at lepton colliders, it 
should be noted that currently no SM $t$-channel diagram is included in 
\pythia8. For example, this would be relevant for studies of Bhabha 
scattering. The SM contribution should therefore be similar to the \pythia8\ 
process, {\tt WeakSingleBoson:ffbar2gmZ}, when only the charged lepton final 
states are considered. A spin-1 unparticle can have chiral couplings to 
fermions which could lead to interesting interference effects \cite{bib:geo07:2}. 
For this reason the chiral properties of the coupling can be specified explicitly 
by two coupling parameters ($g_{XX}$, $g_{XY}$). These can be assigned the 
values 1, 0 or -1 which are multiplied with the different helicity amplitudes, 
$XX ~=~ LL/RR$ and $XY ~=~ LR/RL$. This implies that a vector coupling corresponds 
to, $g_{XX} = g_{XY} = 1$, and an axial coupling to, $g_{XX} = 1, g_{XY} = -1$.

The corresponding graviton processes, where the unparticle is replaced by a 
virtual graviton, are again obtained using the common implementation. This 
has been achieved by the same approach as for the $\unpU/G$ emission processes 
where two constants had to be adjusted. The cross sections given by the spin-2 
unparticle MEs discussed above become identical to the graviton cross section 
expressions in \cite{bib:giu99} when the unparticle parameters are translated 
as follows,
\begin{eqnarray}
d_\unpU = 2 \\
\chi _\unpU = 1 \\
\lambda = \sqrt{4 \pi} \\
\Lambda_\unpU = \Lambda_T
\end{eqnarray}
The virtual graviton exchange processes are UV sensitive, similar to the 
unparticle scenario. The graviton amplitudes are dominated by contributions 
that depend on the UV completion of the theory when the number of large extra 
dimensions\footnote{Which have a similar role to $d_\unpU$ in the unparticle 
model.}, $n>2$. Since this scenario is generally of great interest, the UV 
sensitive part is commonly replaced by an arbitrary cut-off scale, which 
parametrises our ignorance and removes the $n$ dependence. 

\subsection{Treatment of the Effective Theory at High Energies \label{sec:tethe}}
Since the current limits on the effective mass scales related to the different 
processes ($\Lambda_\unpU$, $M_D$ and $\Lambda_T$) are well below the possible 
center--of--mass energies of the hard process, $\hat{s}$, at the LHC, a number 
of different options are available for the case where the hard scale of the process 
approaches or exceeds the scale of validity of the effective theory. These are 
specified by the program parameter {\tt CutOffmode}, which can take the following 
values:
\begin{itemize}
\item{{\tt CutOffmode = 0}: Include all $\hat{s}$ values;}
\item{{\tt CutOffmode = 1}: Suppress the cross section for $\hat{s} > \Lambda_\unpU^2$ ($M_D^2$);}
\item{{\tt CutOffmode = 2}: Gravity form factor, using {\tt SigmaProcess:renormScale2};}
\item{{\tt CutOffmode = 3}: Gravity form factor, using $\mu = E^*_{jet/Z/\gamma}$.}
\end{itemize}
The default option ({\tt CutOffMode = 0}) implies no restrictions, but includes 
the cross section contribution for any $\hat{s}$ value. The first alternative 
option ({\tt CutOffMode = 1}) simply suppresses the cross section~\cite{bib:vac01} 
by a factor $\Lambda_\unpU^4/\hat{s}^2$ at $\hat{s}$ values that exceed the 
mass scale of the effective theory, $\Lambda_\unpU^2$ or $M_D^2$. This truncation 
of the cross section also implies that the $\unpU/G$ mass spectrum is suppressed 
at larger values. The $\unpU/G$ mass spectrum becomes increasingly peaked towards 
larger values with increasing $n$ or $d_\unpU$. For this reason, the suppression 
effect generally becomes more significant for increased values of these parameters. 
In a similar way the effect of this truncation becomes larger with an increased 
transverse energy requirement in the analysis. This option is only implemented 
for the $\unpU/G$ emission processes and further details can be found in \cite{bib:ask09}. 
Starting the truncation directly above the effective mass scale can often be 
considered as conservative, where for example \cite{bib:giu99} estimates the 
effective theory for the LED graviton processes to be perturbative up to about 
$7M_D$. This rather crude method of truncating the cross section was primarily 
intended to be used for verifying that the truncated part, which corresponds to 
the region where the effective theory might not be valid, has a negligible effect 
on the total cross section. However, in the case where it is not negligible the 
truncated value could also be used as a conservative estimate.

The possibility of using a form factor~\cite{bib:hew07} for the gravitational 
coupling has also been implemented,
\begin{equation}
F(\mu ; t, M, n) = \left [ 1 + \left ( \frac{\mu^2}{t^2M^2} \right )^{1+n/2} \right ]^{-1}
\label{eq:ff}
\end{equation}
Here $\mu$ is a renormalisation scale and $t$ is a ${\cal O}(1)$ free parameter 
which relates to the unknown details of the running of the gravitational 
coupling. The parameter $M$ is associated with the cut-off scale of the 
effective theory ($M_D$ or $\Lambda_T$ depending on the process). The form 
factor leads to a weaker gravitational coupling at higher energies and this 
provides a realistic alternative of the behaviour when $\hat{s}$ approaches 
$M^2$. For both real emission and virtual exchange of gravitons the 
renormalisation scale $\mu$ can be specified ({\tt CutOffMode = 2}) to follow 
the choice made by the program parameter, {\tt SigmaProcess:renormScale2}. 
For graviton emission it is possible to set the renormalisation scale equal 
to the $jet/Z/\gamma$ centre--of--mass energy ({\tt CutOffMode = 3}) which 
was used in ~\cite{bib:hew07}. It is argued in \cite{bib:hew07} that the 
form factor can generally prevent virtual graviton exchange amplitudes from 
violating unitarity by requiring $t < 2$, which is suggested as an upper 
bound on this parameter. 


\section{Real Emission, Mono-Jets \label{sec:remj}}
Both the LED graviton and unparticle models can lead to final states 
with a single jet plus missing transverse energy, with balancing 
transverse momenta. The related $G$ emission processes available in 
\pythia8\ consist of {\tt qqbar2Gg, qg2Gq, gg2Gg}, with the program 
parameters {\tt n, MD, CutOffmode} and {\tt t}. The corresponding 
$\unpU$ emission processes are {\tt qqbar2Ug, qg2Uq, gg2Ug}, with the 
relevant parameters {\tt spinU, dU, LambdaU, lambda} and {\tt CutOffmode}. 
The unparticle processes are only available for the spin values 
{\tt spinU = 0,1} and, in both the case of $G$ and $\unpU$ emission, 
the user should match the generated mass distribution to the cross 
section, as discussed in section \ref{sec:iotp}.

This section presents characteristic results related to \Uj\ production 
at the LHC obtained using \pythia8. The processes are validated with 
respect to the results presented in \cite{bib:che07,bib:giu99,bib:riz08}
and, to allow for easy comparison, most of the selection cuts and model 
parameter values have been used consistently with these papers.

\subsection{LED Graviton Emission}
\begin{figure}[tb]
  \centering
  \epsfxsize=0.70\textwidth
  \epsfbox{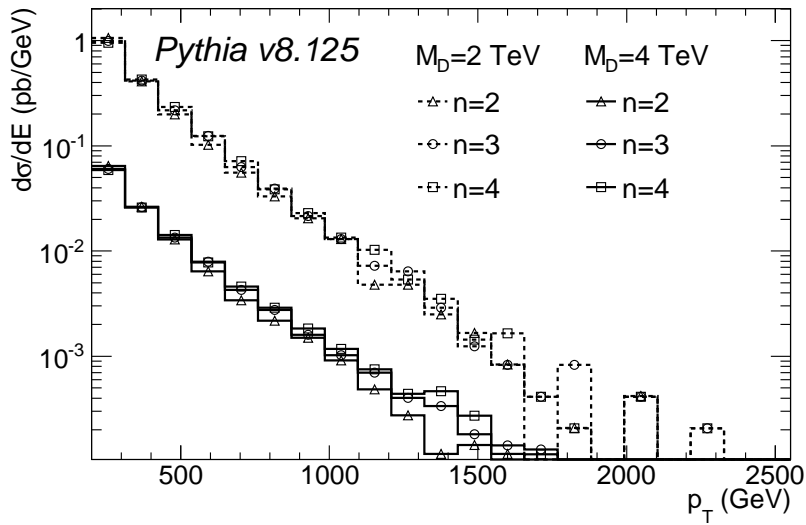}
  \caption{\small Differential cross section of $G + jet$ production at 
    the LHC ($\sqrt{s} = 14$ TeV) as a function of transverse jet momentum. 
    All shapes are normalised to the case $n=2$.}
  \label{fig:ADD_cfr}
\end{figure}

In Fig.~\ref{fig:ADD_cfr} the differential cross section of $G + jet$ 
production at the LHC is shown as a function of the transverse jet momentum 
for different values of $M_D$ and $n$. For these generator level results 
the jet simply refers to the outgoing quark or gluon in the hard interaction. 
The jet has been required to have transverse momentum $p_T>$ 200 GeV together 
with a pseudo-rapidity $|\eta|<$ 2. For the large $\hat{s}$ region the form 
factor ({\it c.f.} Eq.~(\ref{eq:ff})) has been used ({\tt CutOffMode = 2, 
renormScale2 = 1}) with $t = 1$. 
In order to simplify shape comparisons all distributions have been 
normalised to the cross section for $n=2$. As expected, changing one of 
the two parameters has a small effect on the shape of the different 
distributions. The results are consistent, both in terms of shape and 
cross section, with previous studies. The same distributions have also been 
produced using different PDFs: MRST2001, CTEQ6L, GRV94L\footnote{Although 
GRV94L is relatively old it has been included since it was used in some of the 
early graviton papers.}. As expected, the effect of using different PDFs 
is very small.
\begin{figure}[tb]
  \centering
  \epsfxsize=0.70\textwidth
  \epsfbox{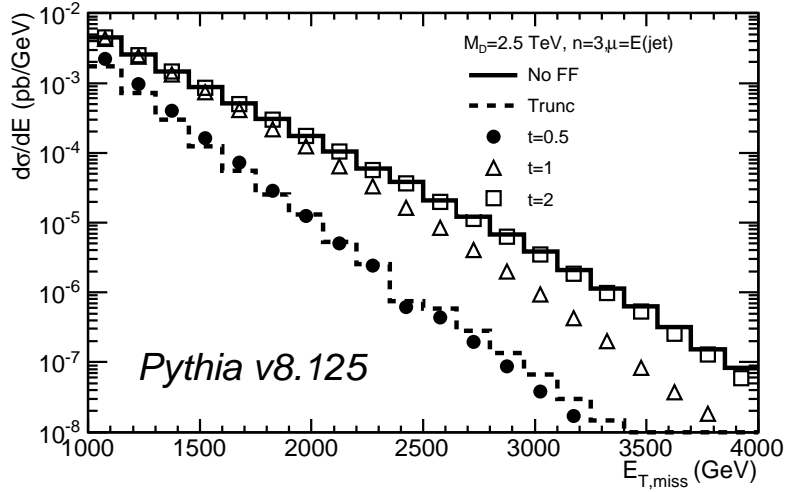}
  \caption{\small Differential $G+jet$ cross section at the LHC 
    ($\sqrt{s} = 14$ TeV) as a function of the missing transverse energy. 
    The plot shows results obtained using the form-factor compared with 
    the case where all $\hat{s}$ values contribute (NoFF) and the case 
    where the cross section is truncated (Trunc).}
  \label{fig:ADD_cfr_FF} 
\end{figure}

Since the high energy collider signals can be sensitive to the large 
$\hat{s}$ region, where the effective theory is not necessarily valid, 
it is interesting to study the effect of the different methods to treat 
this region ({\it c.f.} Sec.~\ref{sec:tethe}). In order to use the form 
factor, the quantity $\mu$ needs to be related to a physical scale in 
the production process and the choice of scale can affect the cross 
sections significantly. For the mono-jet studies presented here, the 
$\mu$ parameter has either been set to the jet energy ({\tt CutOffMode = 3}) 
or alternatively to the jet $p_T$ ({\tt CutOffMode = 2, renormScale2 = 1}).

Fig.~\ref{fig:ADD_cfr_FF} shows the missing transverse energy distributions 
from $G+jet$ production when using the different options related to the 
UV region. The form factor results are compared with the case where all 
$\hat{s}$ values contribute (NoFF) and the case where the cross section 
is truncated (Trunc). The effect of the form factor, which implies 
a softer distribution, is visible together with the expected dependence 
on the model parameters. The effect decreases with increasing $t$, where 
$t=2$ is shown to be similar to the NoFF distribution and the case with 
$t=0.5$ approaches the truncated distribution. In addition, the effect 
increases slightly for $\mu=E({\rm jet})$ compared to $\mu=p_T({\rm jet})$. 
The study showed moderate sensitivity to $n$ and the effect becomes larger 
for smaller values of $M_D$, which is consistent with Eq. (\ref{eq:ff}).

\subsection{Unparticle Emission}
\begin{figure}[tb]
  \centering
  \epsfxsize=0.70\textwidth
  \epsfbox{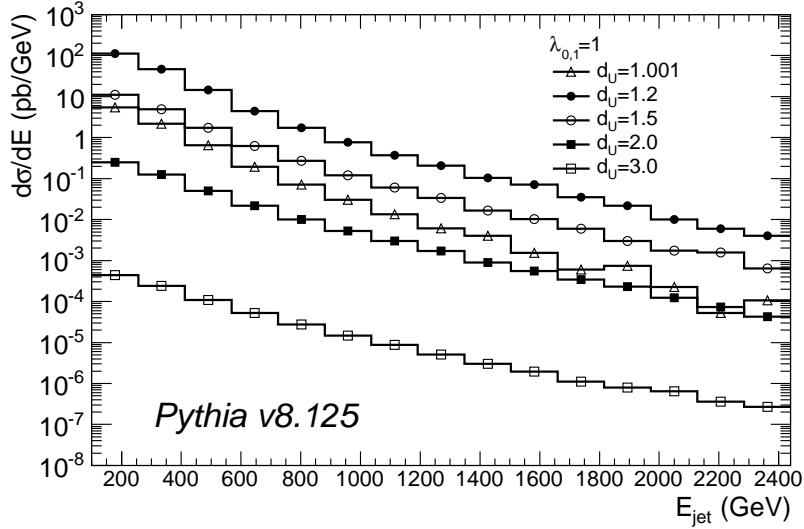}
  \caption{\small Differential cross section of $U+jet$ production at 
    the LHC ($\sqrt{s} = 14$ TeV) as a function of the jet energy. 
    Contributions from spin-0 and spin-1 unparticles are included.}
  \label{fig:U_cfr}
\end{figure}

Fig.~\ref{fig:U_cfr} shows the differential cross section for mono-jet 
production due to \unpU\ emission at the LHC as a function of the jet 
energy. In accordance with~\cite{bib:che07} the cross sections correspond 
to spin-1 unparticle emission through the two processes, \qqUg\ and \qgUq\, 
together with spin-0 unparticle emission due to the gluon process, \ggUg . 
The distributions are shown for a number of $d_\unpU$ values. The parameters 
$\Lambda_\unpU=1$ TeV and $\lambda = 1$ have been used for both the scalar 
and vector processes. Only jets with transverse momenta larger than 100 
GeV and with rapidities $|y(jet)|< 2$ have been considered.
In addition to the cross sections predicted for these unparticle scenarios 
the plot demonstrates the generally small influence of the non-integer 
dimension parameter $d_\unpU$ on the slope. Both the distribution shape 
and cross section values have been found to be consistent with the results 
in the reference literature. 

\begin{figure}[tb]
  \centering
  \epsfxsize=0.70\textwidth
  \epsfbox{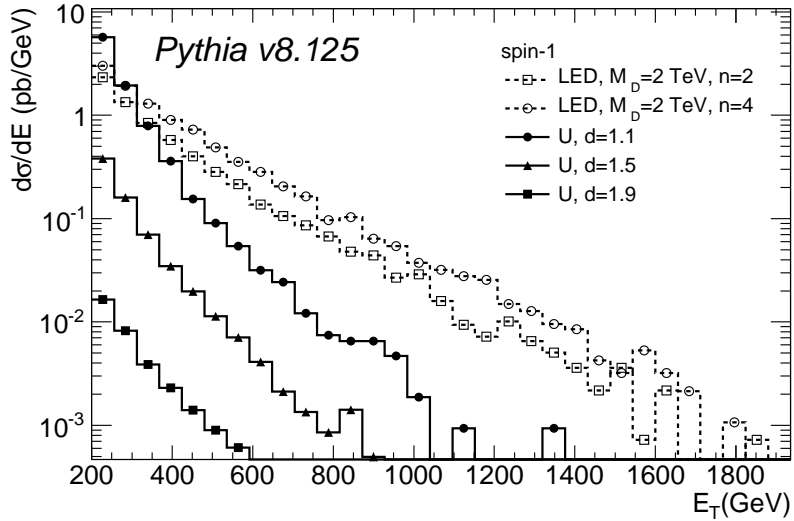}
  \caption{\small Differential cross sections as a function of the transverse 
    jet energy for LED graviton emission and emission of spin-1 unparticles at 
    the LHC ($\sqrt{s} = 14$ TeV).}
  \label{fig:ADD_U_cfr}
\end{figure}

The coherent implementation of the similar $\unpU$ and $G$ processes in 
\pythia8\ simplifies studies of the possibility to distinguish between 
the two models. As an example, a comparison of the differential cross 
section as a function of the transverse jet energy is shown in Fig.~\ref{fig:ADD_U_cfr}. 
The graviton distributions for two values of $n$ are compared to the 
distributions from spin-1 unparticle emission for $d_\unpU =$ 1.1, 1.5 and 
1.9. The unparticle results contain both the two allowed sub-processes and 
the parameters values $\Lambda_\unpU = 1$ TeV and $\lambda = 1$ have been 
used. It is clearly visible that the two models exhibit different spectrum 
slopes almost independently of $n$ or $d_\unpU$ and similar results were found
for the spin-0 unparticle case. 
For the value, $M_D = 2$ TeV, the cross section for LED graviton production 
is significantly higher in the tail than for the unparticle scenarios shown. 
The cross sections, however, scale with $M_D$ or $\Lambda_\unpU$ which allows 
the cross section from either of the two models to be larger or smaller than 
the other. Therefore, as pointed out in \cite{bib:riz08}, it is important to 
also study the shape of the signal in order to understand which new physics 
model is more compatible with a possible mono-jet excess at the LHC.


\section{Virtual Exchange, Photon Pair Production \label{sec:vepp}}
The \pythia8\ processes related to virtual $G$ exchange {\tt ffbar2gammagamma} 
and {\tt gg2gammagamma} are controlled by the program parameters 
{\tt LambdaT, CutOffmode, t}, and for these processes {\tt CutOffmode = 1} 
is not available. The corresponding unparticle processes have the same 
names, but reside in the different unparticle process category as listed 
in the appendix. They are controlled by the following parameters {\tt spinU, 
dU, LambdaU, lambda}, with the two possible spin values {\tt spinU = 0,2}.

This section presents different characteristic results for di-photon 
production due to both virtual $\unpU$ and $G$ exchange. The validation 
of these processes is based on the results presented in \cite{bib:hew07,
bib:che07,bib:kum08,bib:giu99,bib:giu05} and similar selection cuts as well 
as model parameter values have been chosen for easy comparison.

\subsection{Virtual Graviton Exchange}
\begin{figure}[tb]
  \centering
  \epsfxsize=0.70\textwidth
  \epsfbox{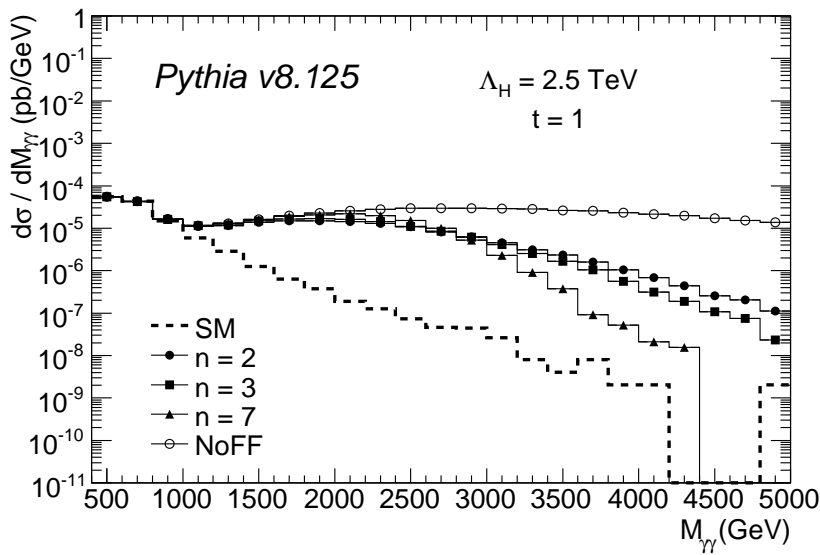}
  \caption{\small 
    The di-photon invariant mass spectrum for LED graviton exchange at the 
    LHC using $\Lambda_{H}=2.5$ TeV. The results are produced with and without 
    using the form factor, assuming various values of $n$.}
  \label{fig:Graviton1}
\end{figure}

Fig.~\ref{fig:Graviton1} shows the invariant mass distribution of the 
two photons produced by LED graviton exchange, with and without using the 
form factor in Eq.~(\ref{eq:ff}). In accordance with~\cite{bib:hew07} the 
mass scale value, $\Lambda_H = 2.5$ TeV, has been used\footnote{The value 
using the $\Lambda_H$ convention was simply translated to, $\Lambda_T = 2.8$ 
TeV, as described in section \ref{sec:pc}.}. The results using the form 
factor are shown for a number of different values of $n$. For virtual 
graviton exchange the renormalisation scale has been set to $\mu^2 = \hat{s}$ 
({\tt CutOffMode = 2, renormScale2 = 4}).
The characteristic effects from the form factor are visible, in particular 
the softening of the invariant mass distribution. As expected from Eq.~(\ref{eq:ff}), 
the distributions decrease faster at large $M_{\gamma\gamma}$ values for large 
$n$ and they intersect at $M_{\gamma\gamma} = \Lambda_T$, which for example in 
Fig.~\ref{fig:Graviton1} corresponds to 2.8 TeV. The study also showed the 
expected behaviour that the graviton contribution becomes significant at larger 
$M_{\gamma\gamma}$ values with increased values of $\Lambda_T$.

\subsection{Virtual Unparticle Exchange}
\begin{figure}[tb]
  \centering
  \epsfxsize=0.70\textwidth
  \epsfbox{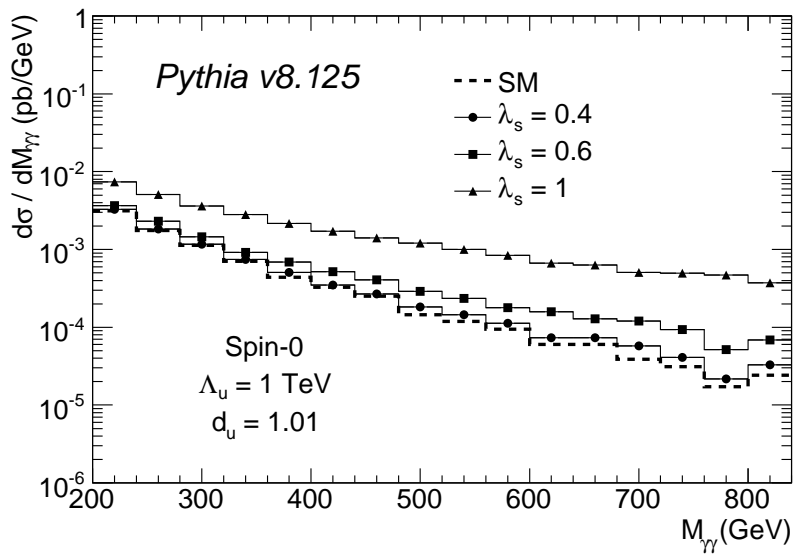}
  \caption{\small Invariant mass distribution of the di-photon system from 
    spin-0 unparticle exchange at the LHC ($\sqrt{s} = 14$ TeV) for different 
    values of the $\lambda_{s}$ and $\lambda_{t}$ couplings.}
  \label{fig:Unparticle1}
\end{figure}

\begin{figure}[tb]
  \centering
  \epsfxsize=0.70\textwidth
  \epsfbox{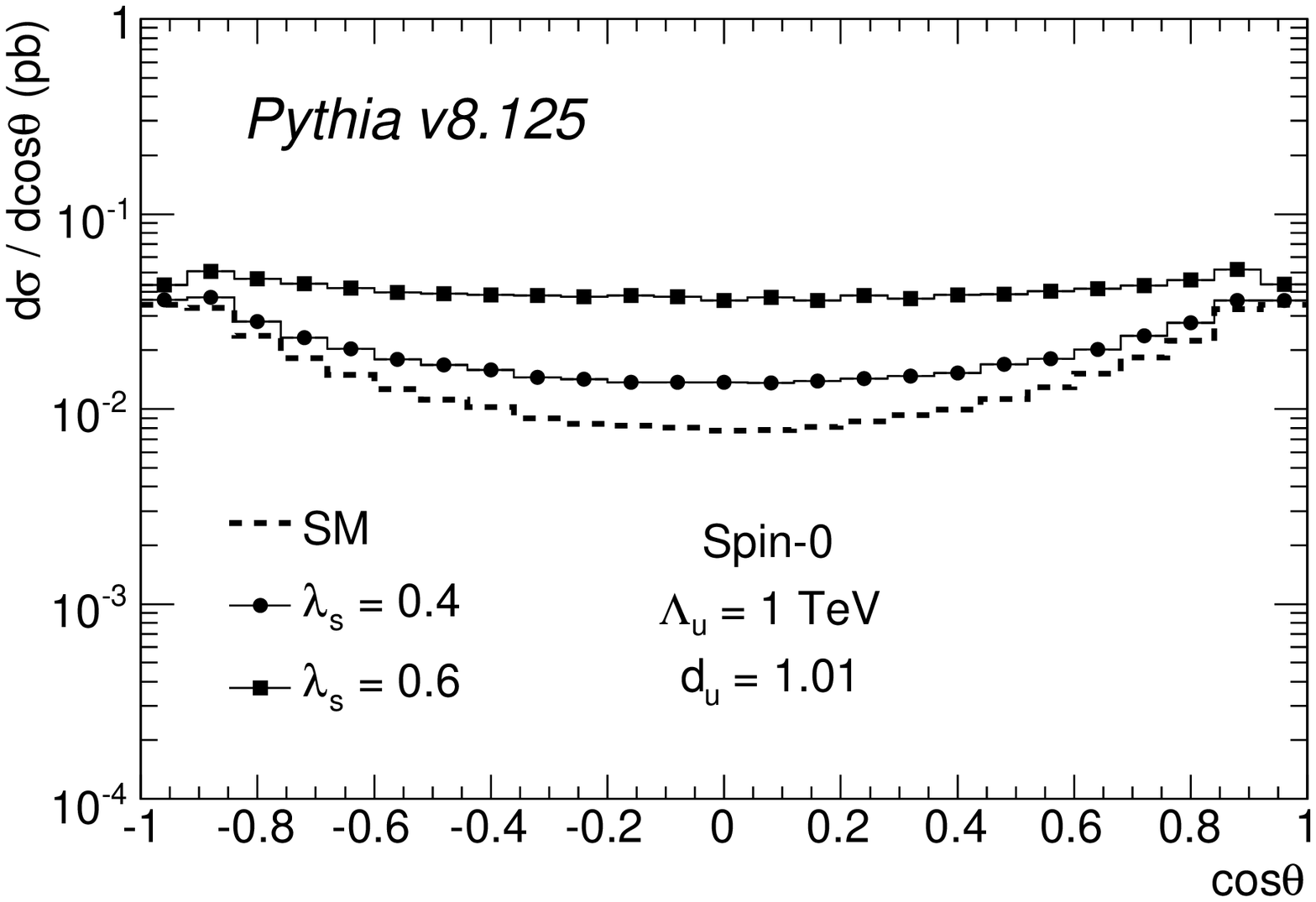}
  \caption{\small Angular distribution, $\cos{\theta ^*_{\gamma}}$, of the 
    photons from spin-0 unparticle exchange at the LHC ($\sqrt{s} = 14$ TeV).}
  \label{fig:Unparticle2}
\end{figure}

The results related to \gg\ production at the LHC from virtual unparticle 
exchange have been compared with~\cite{bib:kum08}. Cuts have been applied 
in all cases on the photon rapidity, $\mid{y^{\gamma}}\mid<2.5$, and transverse 
momentum $p^{\gamma}_{T}>40$ GeV. For the invariant mass distributions the 
cut on the polar angle, $\mid{cos\theta_{\gamma}}\mid<0.8$, has been used 
and for the angular and rapidity distributions only the invariant mass region, 
$600 < M_{\gamma\gamma} < 900$ GeV, has been considered. Fig.~\ref{fig:Unparticle1} 
and \ref{fig:Unparticle2} show the differential cross section as a function 
of $M_{\gamma\gamma}$ and $\cos{\theta _{\gamma}^*}$ respectively, for spin-0 
unparticles, where $\theta _{\gamma}^*$ corresponds to the polar angle in the 
centre--of--mass frame of the photons. 
The invariant mass distribution is shown for three different $\lambda$ 
values and the signal decreases in the expected way with smaller coupling 
constants. The characteristic angular shape of the signal is visible and 
the results are found to be in good agreement with the literature listed 
above. In addition, the effects on various kinematic distributions from 
using different PDFs have been studied by using again the PDF sets MRST2001, 
CTEQ6L, GRV94L and the results showed only small differences.


\section{Virtual Exchange, Lepton Pair Production \label{sec:velp}} 
The available processes in \pythia8\ with virtual graviton exchange leading 
to lepton pairs in the final state (Drell-Yan) correspond to {\tt ffbar2ll} 
and {\tt gg2ll}. The relevant program parameters are {\tt LambdaT, CutOffmode, 
t}, where {\tt CutOffmode = 1} is not available. Similar to the \gg\ case, 
the corresponding unparticle processes have the same names, but reside in 
the different unparticle process category as listed in the appendix. They 
are controlled by the parameters {\tt spinU, dU, LambdaU, lambda, gXX, gXY}, 
with two unparticle spin values available {\tt spinU = 1,2}.

In this section we present a number of results obtained with \pythia8\ 
used to validate the implementation of these processes as well as to 
quantify their characteristic behaviour. The processes are mainly 
validated against the results presented in \cite{bib:hew07,bib:che07,
bib:giu99,bib:giu05} and the choices of model parameter values and several 
selection criteria have also been based on these papers. We mainly focus 
the study on lepton pair production at the LHC. However, since most of the 
literature related to effects from chiral spin-1 unparticle couplings is 
available for $e^+e^-$-collisions we also present some validation results 
in that context.

\subsection{Virtual Graviton Exchange}
\begin{figure}[tb]
  \centering
  \epsfxsize=0.70\textwidth
  \epsfbox{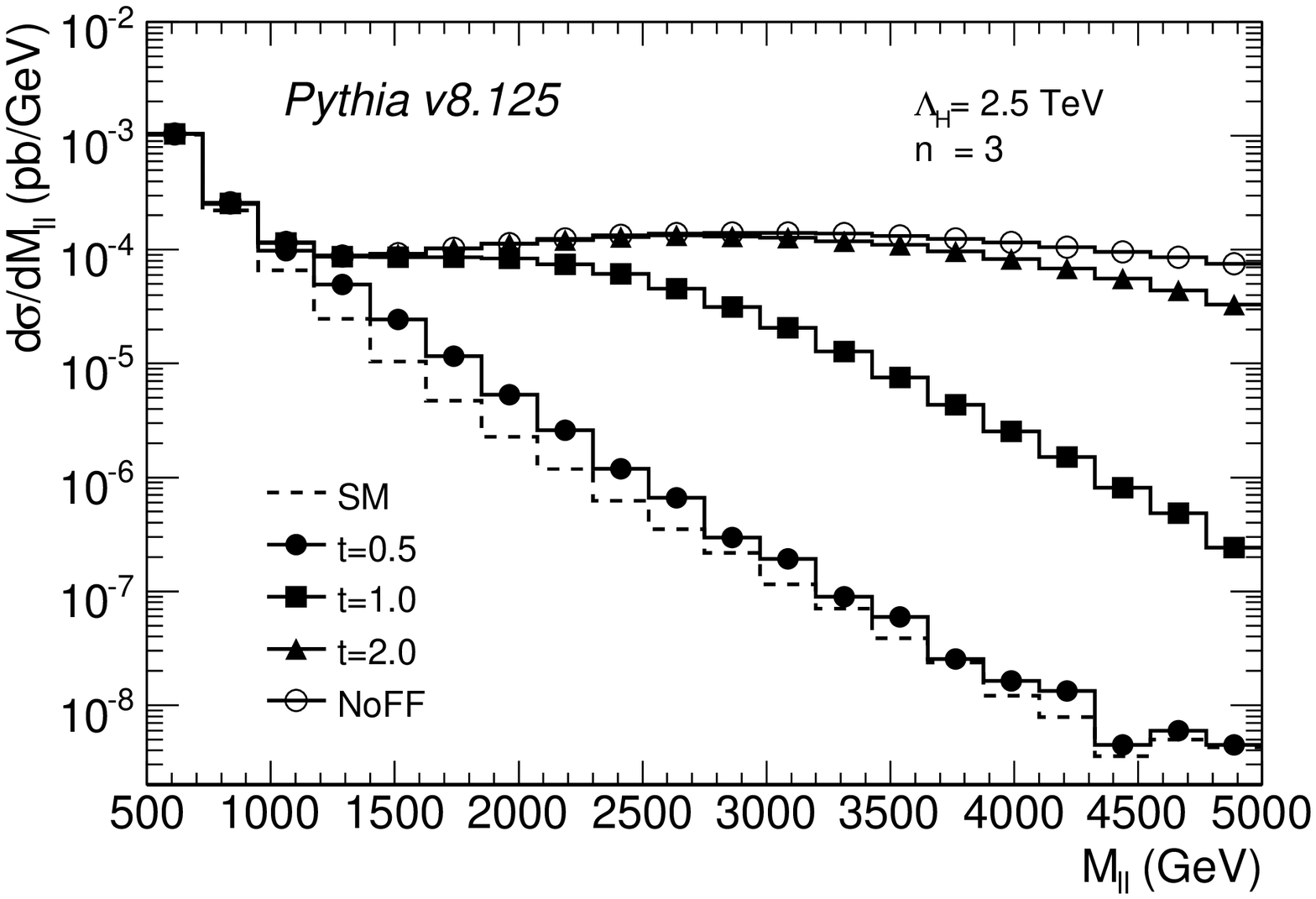}
  \caption{\small Differential cross section for the production of two
    charged leptons at the LHC as a function of their invariant mass. 
    The results are shown for various LED scenarios with $n=3$ and 
    $\Lambda_{H}=2.5$~TeV.}
  \label{fig:LED_ll}
\end{figure}

In addition to the di-photon production at the LHC, discussed in Sec. 
\ref{sec:vepp}, the production of two leptons can be influenced by the 
effects of virtual graviton exchange. For this reason, the invariant mass 
distribution of the two final-state leptons has been studied for a large 
number of different parameter settings of the LED model and agreement with 
the reference literature has been found. As an example, Fig.~\ref{fig:LED_ll} 
shows the differential cross section as a function of the invariant mass 
of the lepton pair for the cut-off scale, $\Lambda_{H}=2.5$~TeV, in scenarios 
with $n=3$. As for the \gg\ process the mass scale value in the common 
$\Lambda_H$ convention is used. In contrast to the \gg\ case the results 
are given for a number of different $t$ parameter values at a fixed $n$. 
The results show the strongly $t$-dependent increase of the cross section 
at high masses for the LED scenarios with respect to the SM. As expected 
from Eq.~(\ref{eq:ff}), the differential cross section converges towards 
the SM prediction for decreasing values of $t$, while for larger $t$ values, 
{\it e.g.}~$t=2.0$, it approaches the LED scenarios where the form factor 
is not used (NoFF). In addition, the dependence of the cross section on the 
number of extra dimensions, $n$, has been studied and a similar behaviour to 
the di-photon production ({\it c.f.}~Fig.~\ref{fig:Graviton1}) has been found 
as well as a good agreement with the results in the literature.

\subsection{Virtual Unparticle Exchange}
Similar to the virtual graviton exchange discussed above, the virtual
exchange of unparticles can modify the invariant mass spectrum and the
angular distribution of final-state lepton pairs at the LHC as well as 
in $e^+ e^-$-collisions. The \pythia8\ results from various settings 
of the unparticle model parameters have been compared to the original 
literature and again good agreement has been found with the results
for $p\bar{p}$ collisions at the Tevatron as presented in \cite{bib:che07}.

\begin{figure}[tb]
  \centering
  \epsfxsize=0.70\textwidth
  \epsfbox{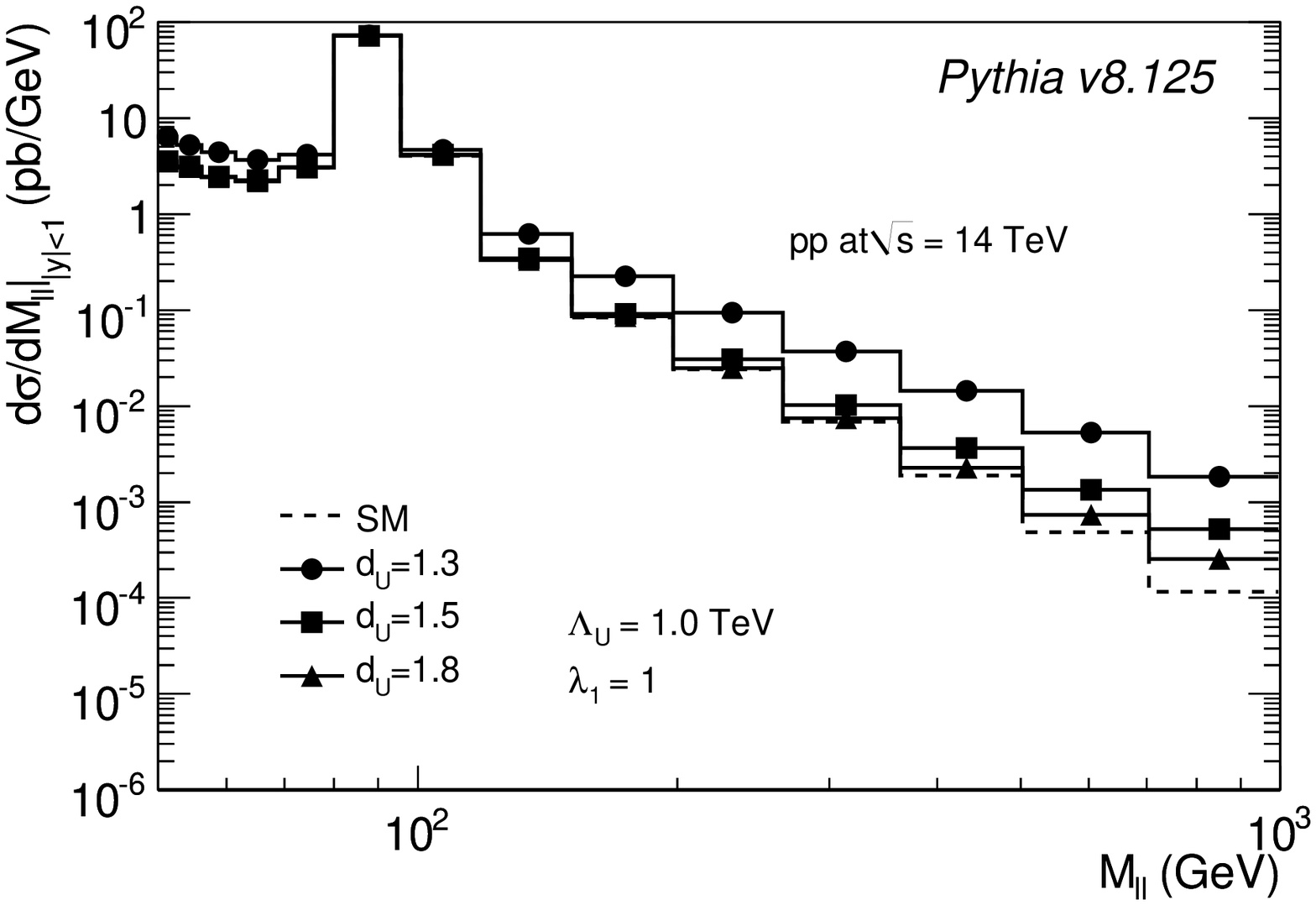}
  \caption{\small Differential cross section for the production of two
    charged leptons as a function of their invariant mass. The results 
    are shown for scenarios with unparticles using different values of 
    $d_\unpU$ at the LHC.}
  \label{fig:Unp_ll}
\end{figure}

Fig.~\ref{fig:Unp_ll} shows the differential Drell-Yan cross section 
at the LHC as a function of the invariant mass of the lepton pair for 
various values of $d_\unpU$, assuming unparticles with spin-1 and 
vector-like 4-fermion interactions. A cut on the lepton pair rapidity 
($|y|<1$) has been applied, which ensures that only central leptons are 
selected. As expected, the characteristic increase of the differential 
cross section due to the unparticle signal is more prominent for small 
values of $d_\unpU$, {\it e.g.}~$d_\unpU=1.3$, both above and below the 
$Z$ pole.

\begin{figure}[tb]
  \centering
  \epsfxsize=0.49\textwidth
  \epsfbox{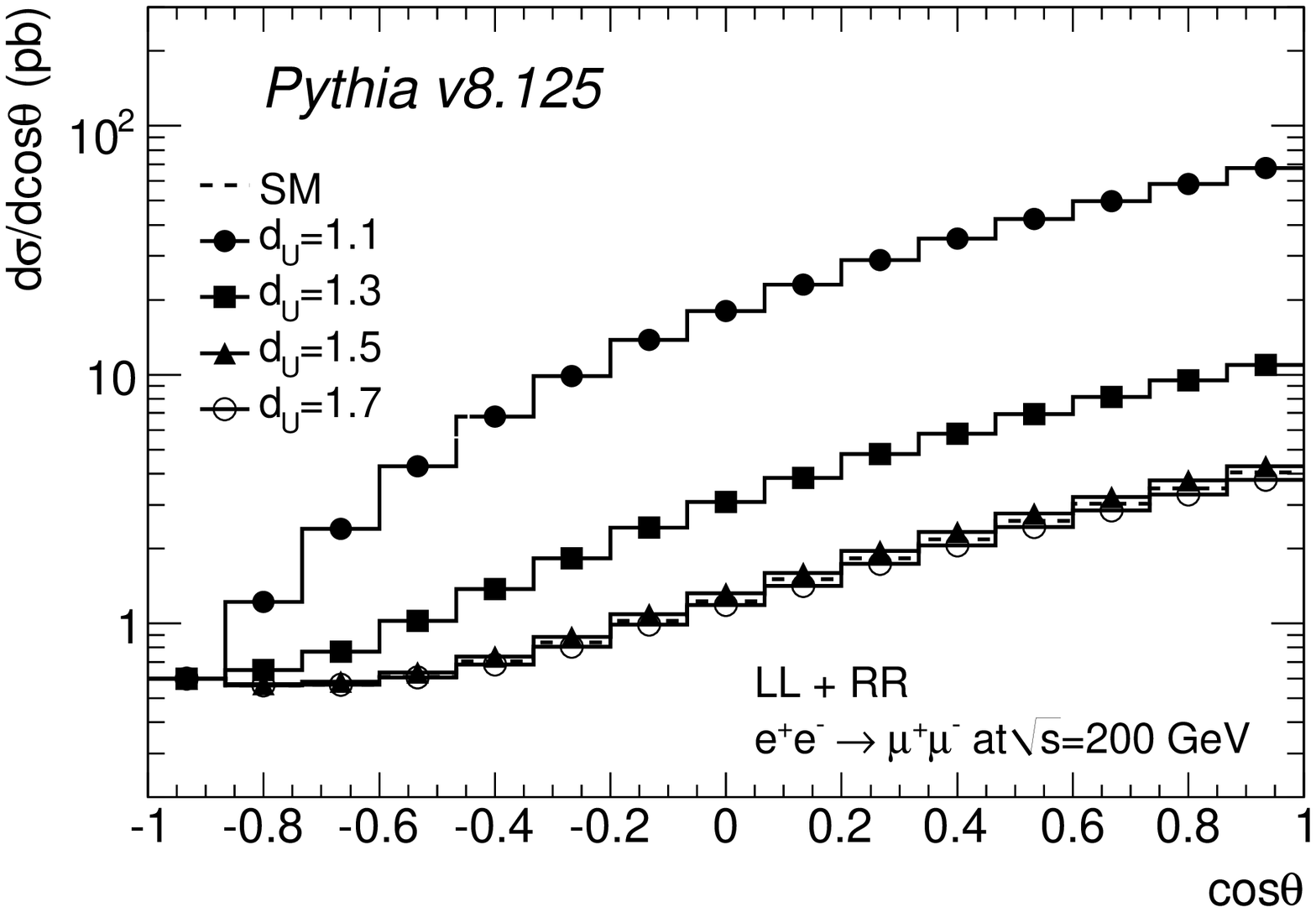}
  \epsfxsize=0.49\textwidth
  \epsfbox{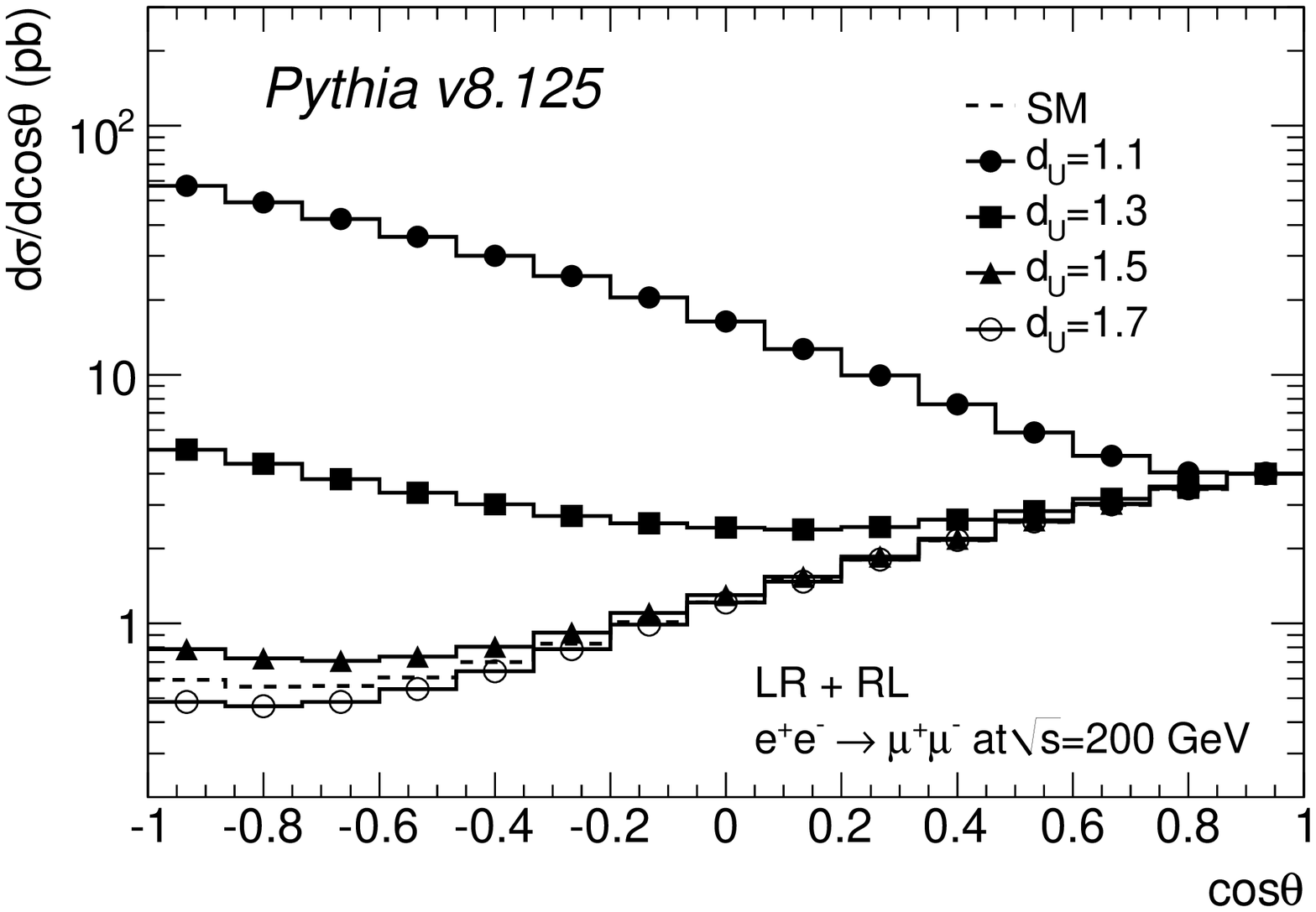}
  \caption{\small Angular distribution for $e^+e^- \rightarrow \mu^+\mu^-$ 
    including unparticle exchange, for various values of $d_\unpU$ at
    $\sqrt{s}=200$~GeV, assuming  $LL+RR$ (left) and $LR+RL$ (right) 4-fermion 
    interactions. }
  \label{fig:Unp_cosTh}
\end{figure}

The possibility of chiral spin-1 unparticle couplings could imply various 
effects, both affecting the invariant mass as well as the angular 
distribution of the final-state lepton pair. These effects have mainly been 
studied for $e^+e^-$-collisions in the literature and we therefore include 
the angular differential cross section obtained by \pythia8\ for the SM 
and unparticle model, using various $d_\unpU$ values, in $e^+e^-$-collisions 
with $\sqrt{s}=200$~GeV. Fig.~\ref{fig:Unp_cosTh} shows the angular 
distributions for $LL+RR$ (left) and $LR+RL$ (right) interactions, where 
the strong impact of a chiral coupling on the angular distribution is 
demonstrated. The parameter values $\Lambda_\unpU = 1$ TeV and $\lambda = 1$ 
have been used. The angle $\theta$ is defined as the emission angle of 
the $\mu^{-}$ relative to the momentum vector of the incoming electron. 

The angular dependence of the differential cross section can be investigated 
by studying the forward-backward asymmetry which is commonly defined as
\begin{equation}
A_{FB} = \frac{N_F-N_B}{N_F+N_B}
\end{equation}
where $N_F$ and $N_B$ are the numbers of forward ($\cos\theta>0$) and 
backward ($\cos\theta<0$) produced events, respectively.  At hadron 
colliders the determination of the angle $\theta$ is more complicated, 
since the direction of the incoming fermion is unknown. However,  $A_{FB}$ 
is foreseen to be studied at the LHC~\cite{bib:atlas_csc} exploiting the 
fact that in annihilations of a valence quark and sea anti-quark the 
valence quark has, on average, a larger momentum than the sea quark. The 
resulting longitudinal momentum of the di-lepton system approximates the 
quark direction and, therefore, the angle $\theta$ between the lepton 
and quark in the di-lepton rest frame can be calculated. For the results 
presented here we use this method at generator level. In addition, we 
require, $|y| > 1$, for the lepton rapidity and that at least one electron 
is in the central region, $|\eta|<2.5$. 

\begin{figure}[tb]
  \centering
  \epsfxsize=0.49\textwidth
  \epsfbox{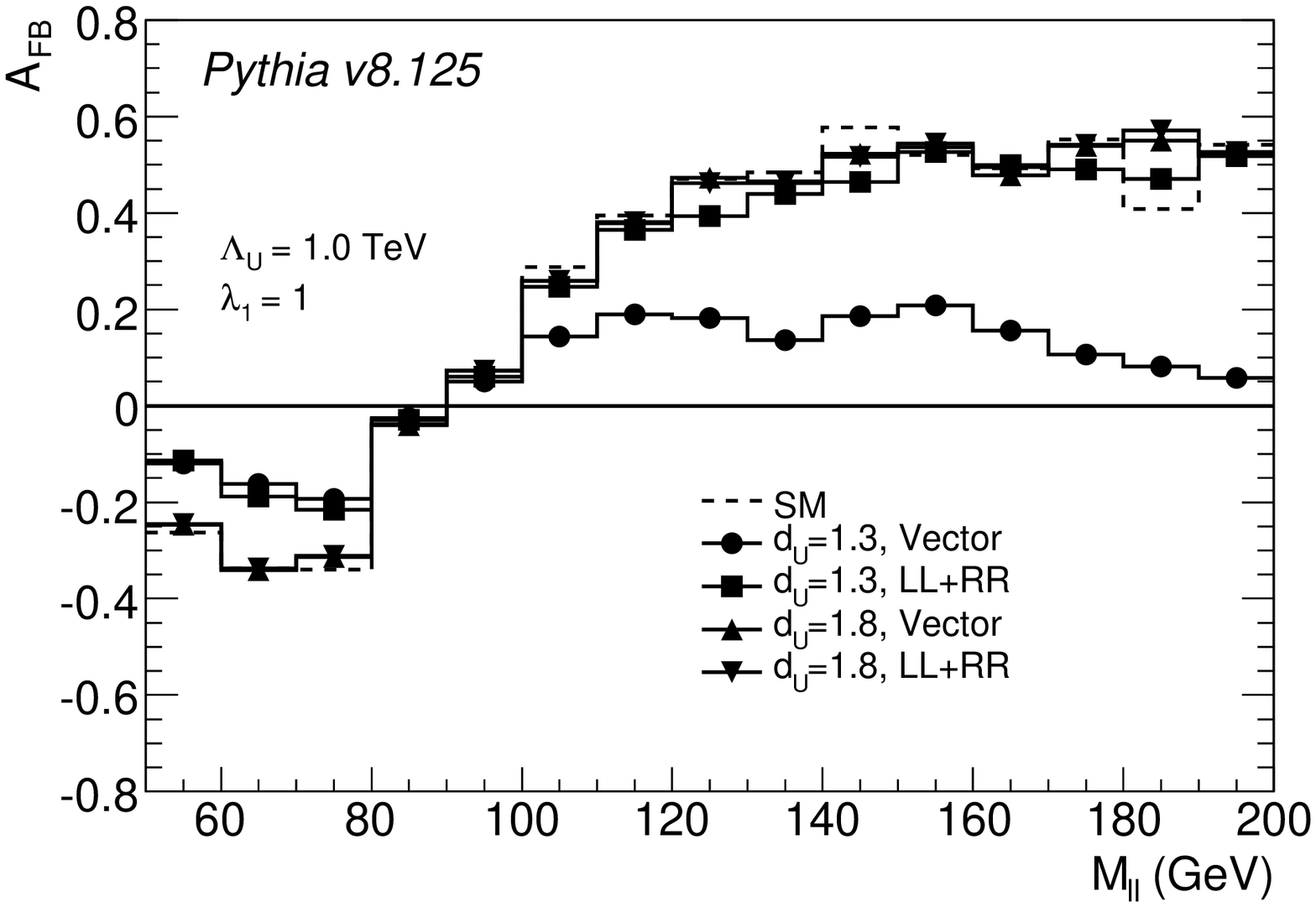}
  \epsfxsize=0.49\textwidth
  \epsfbox{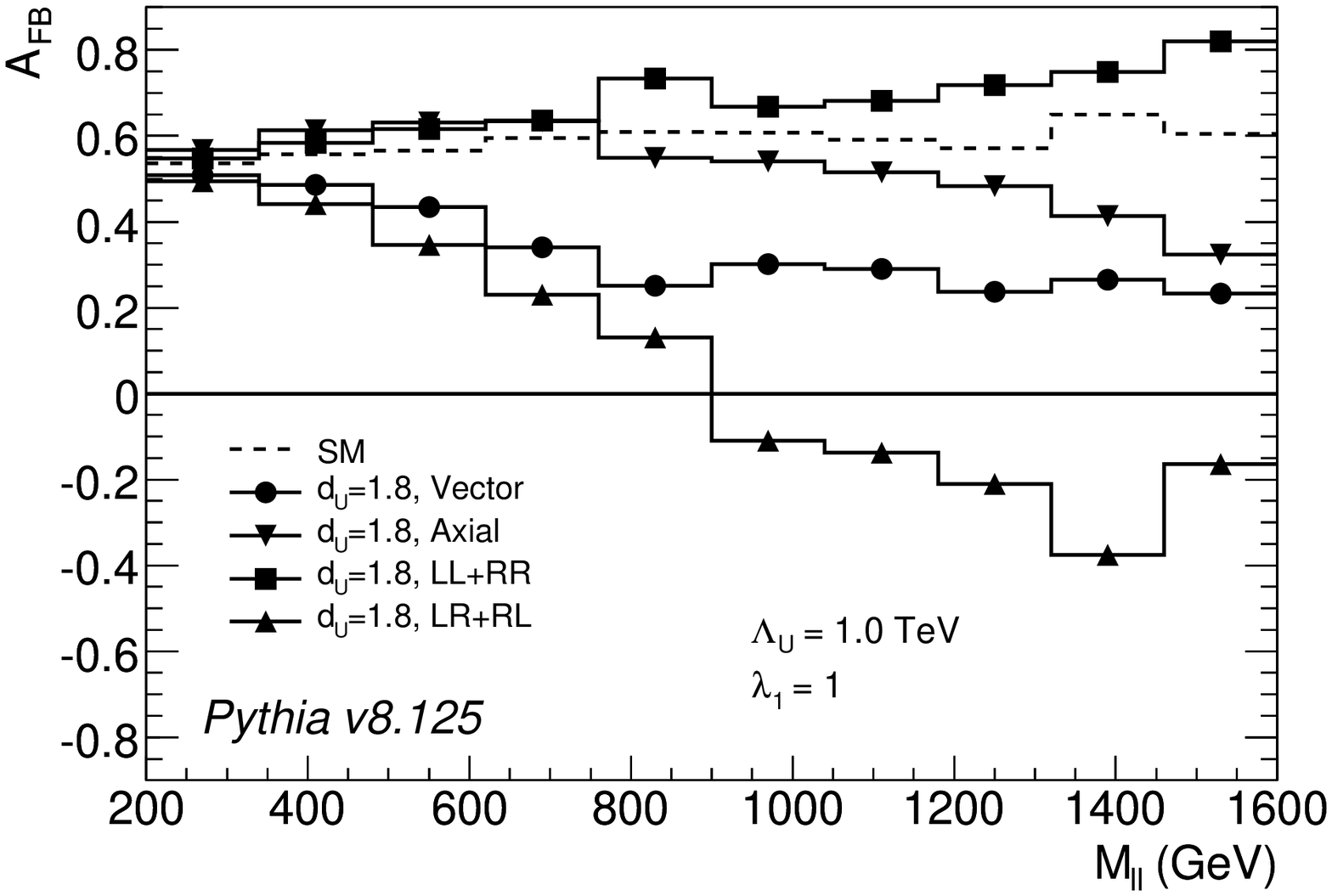}
  \caption{\small $A_{FB}$ as a function of the invariant mass of the 
    lepton pair at the LHC. The results are shown for $d_\unpU=1.3,
    1.8$ with vector-like and $LL+RR$ unparticle interactions (left) 
    as well as for $d_\unpU=1.8$ with various chiral interactions (right). 
    In both figures spin-1 unparticle exchange is assumed.}
  \label{fig:Unp_AFB}
\end{figure}

In Fig.~\ref{fig:Unp_AFB} the estimate of $A_{FB}$, using the method 
described above, is shown for the first time together with effects 
from unparticle exchange. The two plots show the invariant mass range 
50 to 200 GeV (left) and 200 to 1600 GeV (right) respectively for 
various unparticle scenarios and the SM. It can been seen that certain 
unparticle scenarios ({\it e.g.} $d_\unpU = 1.3$) lead to large deviation 
from the SM even at mass values below the $Z$ pole which are already 
investigated in detail by past experiments~\cite{bib:afbref}. However, 
scenarios ({\it e.g.} $d_\unpU = 1.8$) resulting in sizable effects 
only at larger mass values also exist, {\it i.e.} they are consistent 
with the SM in the already studied region at smaller masses.

Experimental effects, in particular statistics, will pose a significant 
challenge for the reconstruction of $A_{FB}$. A detailed experimental 
study is therefore needed to estimate the feasibility. However, in the 
case of an observation of a signal in the high mass tail ({\it c.f.}~Fig.~\ref{fig:LED_ll}), 
a study of $A_{FB}$ as a function of the invariant mass could be of 
interest in order to confirm a non-SM content in the Drell-Yan spectrum.


\section{Full Event Simulation}

In order to allow for easy comparison with results in the literature as 
well as to clearly illustrate the effects related to the model parameters, 
the results from the various processes presented in this paper have been
considered at the level of the hard process.

The processes are, however, integrated within the \pythia8\ framework 
of parton showers, underlying event activity, hadronisation and unstable 
particle decays. These steps both increase the complexity of the events 
as well as affect their kinematic properties and are normally required 
for proper comparisons between simulated phenomena and experimental 
data.

\begin{figure}[tb]
  \centering
  \epsfxsize=0.70\textwidth
  \epsfbox{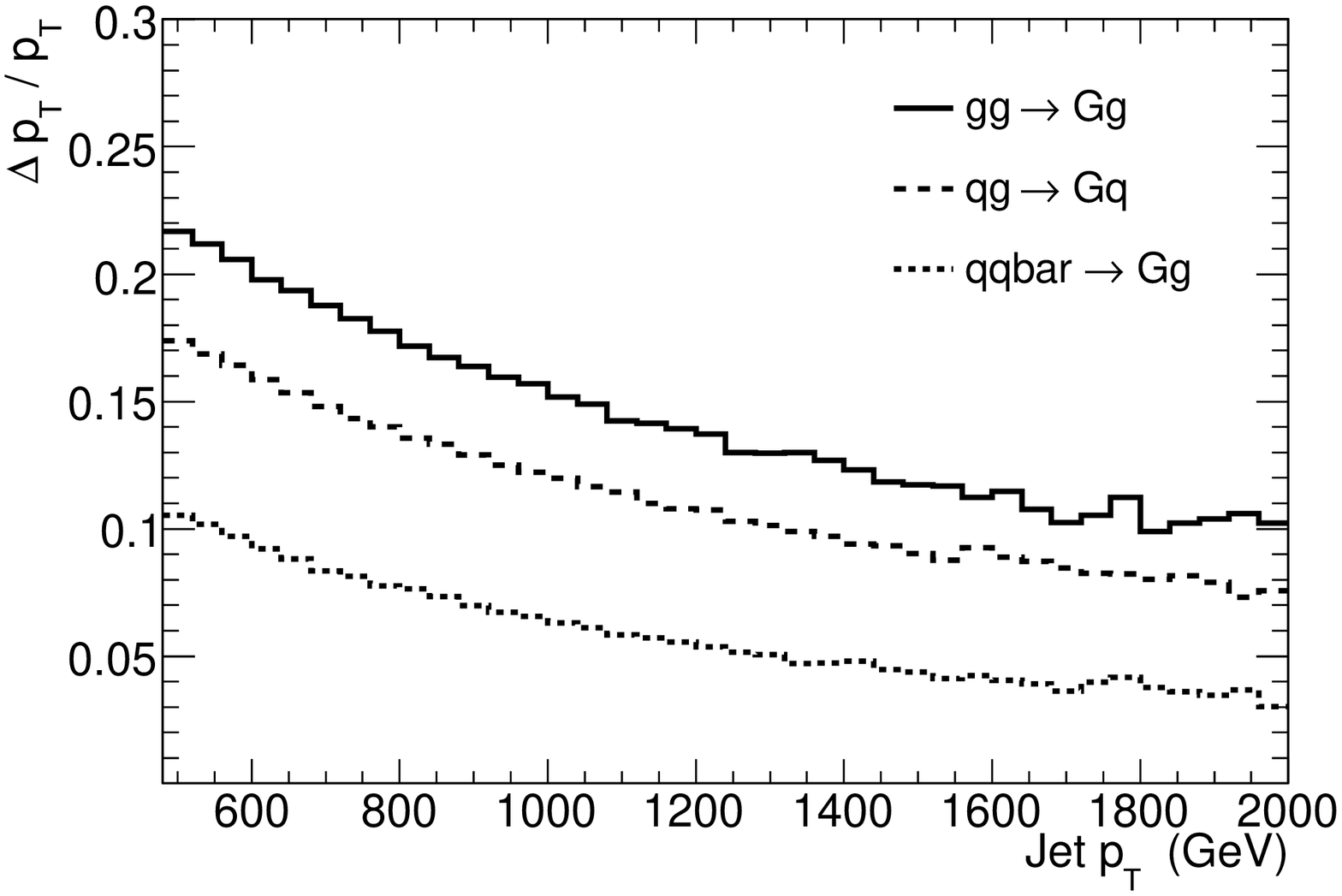}
  \caption{\small Relative $p_{T}$ difference between the graviton and jet 
    due to initial state radiation. The results are shown for $G + jet$ 
    production with $n=6$ and $M_{D}=4$~TeV.}
  \label{fig:ISR}
\end{figure}

As an example, Fig~\ref{fig:ISR} shows how initial state radiation (ISR) 
affects the $p_T$ in $G + jet$ production.
At leading order the $G$ and $jet$ have exactly balanced $p_T$, but 
additional radiation from ISR, or other higher order effects, breaks 
this balance. Fig~\ref{fig:ISR} shows the $p_{T}$ difference between the 
$G$ and the $jet$ after ISR, relative to the original jet $p_T$ in the 
hard process. This is shown as a function of the jet $p_T$ at the hard 
process level and for the different $G + jet$ sub-processes separately.
As seen in the plot, ISR affects the sub-processes noticeably, however, 
to different extent and the imbalance decrease for higher $p_T$ jets. 
This clearly illustrates the importance of building the processes into the 
simulation of the full event structure in Pythia in order to model 
such effects.


\section{Conclusions}
Both large extra dimensions as well as unparticle models have become very 
popular among possible extensions of the SM. Many of the main processes 
related to LED searches have analogous processes in the unparticle scenario, 
which are often more general in terms of model parameters. A common way 
to implement the similar processes has been found within the Monte Carlo 
generator \pythia8, where only a few translations of model dependent 
constants are necessary. In addition to technical benefits, the common 
implementation makes comparisons between the processes from the two models 
consistent and simplifies studies of how differences at the process level 
for a certain final state can affect the measured observables. The processes 
can be used together with several options for the treatment of the UV region 
of the effective theory. In this paper, the options related to the individual 
processes have been discussed together with results obtained by \pythia8. 
In addition to providing references for the \pythia8\ program, the results 
illustrate characteristic effects of the model parameters and various key 
features of the different kinematic distributions. The results presented 
in this paper have been focused on mono-jet, di-photon and di-lepton final 
states at the LHC.


\section{Acknowledgment}
SA would like to express special thanks to Wei-Yee Keung (UIC) for 
providing the formulae related to the scalar $Z\unpU$ process and to 
Henrik Jansen (RWTH) for related helpful feedback. This work was funded 
in the UK by STFC and by the German Research Foundation (DFG) in the 
Collaborative Research Centre (SFB) 676 "Particle, Strings and the Early 
Universe" located in Hamburg. This work was supported in part by the EU 
Marie Curie Research Training Network ``MCnet'', under contract number 
MRTN-CT-2006-035606.

\appendix

\section*{Appendix - $\unpU / G$ Specific Parameters in \pythia8\ }

\noindent In the \pythia8\ program the $\unpU / G$ processes are turned 
on/off by the following parameters, \\
\begin{tabbing}
\={\tt ExtraDimensionsUnpart:ffbar2Ugamma}     \hspace*{1.5cm}\=-- Real $\unpU$ emission, $f\bar{f} \rightarrow \unpU + \gamma$. \\
\>{\tt ExtraDimensionsUnpart:ffbar2UZ}         \>-- Real $\unpU$ emission, $f\bar{f} \rightarrow \unpU + Z$. \\
\>{\tt ExtraDimensionsUnpart:qqbar2Ug}         \>-- Real $\unpU$ emission, $q\bar{q} \rightarrow \unpU + g$. \\
\>{\tt ExtraDimensionsUnpart:qg2Uq}            \>-- Real $\unpU$ emission, $qg \rightarrow \unpU + q$. \\
\>{\tt ExtraDimensionsUnpart:gg2Ug}            \>-- Real $\unpU$ emission, $gg \rightarrow \unpU + g$. \\
\>{\tt ExtraDimensionsLED:ffbar2Ggamma}        \>-- Real $G$ emission, $f\bar{f} \rightarrow G + \gamma$. \\
\>{\tt ExtraDimensionsLED:ffbar2GZ}            \>-- Real $G$ emission, $f\bar{f} \rightarrow G + Z$. \\
\>{\tt ExtraDimensionsLED:qqbar2Gg}            \>-- Real $G$ emission, $q\bar{q} \rightarrow G + g$. \\
\>{\tt ExtraDimensionsLED:qg2Gq}               \>-- Real $G$ emission, $qg \rightarrow G + q$. \\
\>{\tt ExtraDimensionsLED:gg2Gg}               \>-- Real $G$ emission, $gg \rightarrow G + g$. \\
\>{\tt ExtraDimensionsUnpart:ffbar2gammagamma} \> -- Virtual $\unpU$ exchange, $f\bar{f} \rightarrow (\unpU ^*) \rightarrow \gamma \gamma$. \\
\>{\tt ExtraDimensionsUnpart:gg2gammagamma}    \> -- Virtual $\unpU$ exchange, $gg \rightarrow (\unpU ^*) \rightarrow \gamma \gamma$. \\
\>{\tt ExtraDimensionsUnpart:ffbar2ll}         \> -- Virtual $\unpU$ exchange, $f\bar{f} \rightarrow (\unpU ^*) \rightarrow \ell \bar{\ell}$. \\
\>{\tt ExtraDimensionsUnpart:gg2ll}            \> -- Virtual $\unpU$ exchange, $gg \rightarrow (\unpU ^*) \rightarrow \ell \bar{\ell}$. \\
\>{\tt ExtraDimensionsLED:ffbar2gammagamma}    \> -- Virtual $G$ exchange, $f\bar{f} \rightarrow (G^*) \rightarrow \gamma \gamma$. \\
\>{\tt ExtraDimensionsLED:gg2gammagamma}       \> -- Virtual $G$ exchange, $gg \rightarrow (G^*) \rightarrow \gamma \gamma$. \\
\>{\tt ExtraDimensionsLED:ffbar2ll}            \> -- Virtual $G$ exchange, $f\bar{f} \rightarrow (G^*) \rightarrow \ell \bar{\ell}$. \\
\>{\tt ExtraDimensionsLED:gg2ll}               \> -- Virtual $G$ exchange, $gg \rightarrow (G^*) \rightarrow \ell \bar{\ell}$. \\
\end{tabbing}
\noindent The model parameters related to the LED graviton processes are 
specified by,
\begin{tabbing}
\={\tt ExtraDimensionsLED:n}          \hspace*{2.0cm} \=-- \=Number of large extra dimensions, $n$. \\
\>{\tt ExtraDimensionsLED:MD}         \>-- Fundamental scale of $D$ dimensional gravity, $M_D$. \\ 
\>{\tt ExtraDimensionsLED:LambdaT}    \>-- Cut-off scale for virtual $G$ exchange, $\Lambda_T$. \\
\>{\tt ExtraDimensionsLED:CutOffmode} \>-- This parameter specifies the treatment of the high \\
\>\>\>energy contributions. Possible values are 0 to 3. \\ 
\>{\tt ExtraDimensionsLED:t}          \>-- Form factor parameter, $t$. \\
\end{tabbing}
\noindent The model parameters related to the unparticle processes are 
specified by,
\begin{tabbing}
\={\tt ExtraDimensionsUnpart:spinU} \hspace*{1.1cm} \=-- \=The unparticle spin. Possible values are, 0, 1 or 2. \\
\>{\tt ExtraDimensionsUnpart:dU}         \>-- Scale dimension parameter, $d_\unpU$. \\ 
\>{\tt ExtraDimensionsUnpart:LambdaU}    \>-- Renormalization scale, $\Lambda_\unpU$. \\ 
\>{\tt ExtraDimensionsUnpart:lambda}     \>-- Coupling to the SM fields, $\lambda$. \\ 
\>{\tt ExtraDimensionsUnpart:gXX}        \>-- Options for the LL/RR helicity contributions, $g_{XX}$. \\
\>\>\>This parameter is only relevant for the spin-1 \ll\ \\
\>\>\>process, where the available options 0, 1 and 2 \\
\>\>\>correspond to the helicity amplitudes being \\
\>\>\>multiplied by a factor 1, 0 or -1. \\ 
\>{\tt ExtraDimensionsUnpart:gXY}        \>-- Options for the LR/RL helicity contributions, $g_{XY}$. \\
\>\>\>This parameter is only relevant for the spin-1 \ll\ \\
\>\>\>process, where the available options 0, 1 and 2 \\ 
\>\>\>correspond to the helicity amplitudes being \\
\>\>\>multiplied by a factor 1, 0 or -1. \\ 
\>{\tt ExtraDimensionsUnpart:CutOffmode} \>-- This parameter specifies the treatment of the high \\ 
\>\>\>energy contributions. Possible values are 0 or 1. \\ 
\end{tabbing}
\noindent More details on the meaning of the model parameters are given in the text.

\section*{Appendix - Scalar Unparticle Production, $f \bar{f} \rightarrow Z \unpU $.}

The relevant scalar unparticle interaction terms for this process are 
given by,
\begin{equation} 
{\cal L} \supset
{O_{\cal U} \over \Lambda_{\cal U}^{d_{\cal U}-1}} 
\bar f (\lambda_0 \mbox{\boldmath{1}} +\lambda_0'i\gamma_5) f 
\end{equation}
Due to CP conservation, the two couplings $\lambda_0$ or $\lambda'_0$ 
are only considered separately. Since the matrix element expression 
of the two cases becomes identical in the absence of any interference 
with the SM, only one common $\lambda$ value is used for the 
implementation in accordance with Eq. (\ref{eq:unpscalar}). These 
couplings are usually suppressed by a quark mass factor, because of 
the chirality flipping, and are therefore expected to be small in the 
scalar unparticle case. However, in the phenomenological spirit of 
this paper the choice of restricting model parameter values is left 
to the user.

As described in \cite{bib:ask09}, the cross section for the, 
$f \bar{f} \rightarrow Z{\cal U}$, process is given by,
\begin{equation}
\frac{d^2\sigma}{dP_{\unpU}^2dt} = \frac{|\bar{M}|^2}{16 \pi \hat{s}^2} 
\frac{A_{d_{\unpU}}}{2 \pi \Lambda _{\unpU}^2} 
\left( \frac{dP_{\unpU}^2}{\Lambda_{\unpU}^2} \right) ^{d_{\unpU} - 2}
\end{equation}
where $|\bar{M}|$ is the colour and spin averaged matrix element and 
$P_\unpU^2$ is the invariant mass of the unparticle, $0 \le P_\unpU^2 \le (\sqrt{s}-m_Z)^2$.
The matrix element is given by \cite{bib:keu09},
\begin{eqnarray} 
|\bar{M}|^2= \frac{1}{4 N_c}\left( {e^2\over\sin^2\theta_W\cos^2\theta_W}\right)
 (g_{L,q}^2+g_{R,q}^2) \lambda^2 |A|^2 \\
|A|^2 =4\left[
-{s\over t}-\left(1-{m_Z^2\over t}\right)\left(1-{P_\unpU^2\over t}\right)
-{s\over u} \right. \\
\qquad \left.
   -\left(1-{m_Z^2\over u}\right)\left(1-{P_\unpU^2\over u}\right)
   +2\left(1-{P_\unpU^2\over t}\right)\left(1-{P_\unpU^2\over u}\right) \right]
\end{eqnarray}
with 
$g_{L,d}=-{1\over2}+{1\over3} \sin^2\theta_W$, $g_{R,d}=+{1\over3} \sin^2\theta_W$, 
$g_{L,u}={1\over2}-{2\over3} \sin^2\theta_W$ and $g_{R,u}=-{2\over3} \sin^2\theta_W$.
Regarding the amplitude formula, $|A|^2$, the first two terms inside 
the square brackets result from the $t$-channel, the following two terms 
from the $u$-channel and the last term from the interference.

In the same way as for the other spin versions of this process, the 
process where a photon is emitted together with an unparticle has been 
obtained by the following photon limit of the $Z$ matrix element,
\begin{eqnarray}
m_Z \rightarrow 0  \\
\frac{g_L^2 + g_R^2}{2} \rightarrow Q^2 \\
\frac{1}{\sin{\theta_W}\cos{\theta_W}} \rightarrow 1
\end{eqnarray}
where $Q$ is the electric charge of the incoming fermions.

\clearpage


\end{document}